\pdfoutput=1

\documentclass[11pt,twoside,a4paper,cmspaper,final,collab]{cms-tdr}
\def\svnVersion{461479P}\def\cmsCernNoTag{CERN-EP-2018-011}\def\cmsCernDate{\today}\def\cmsMessage{Published in the Journal of High Energy Physics as \href{http://dx.doi.org/10.1007/JHEP05(2018)006}{\doi{10.1007/JHEP05(2018)006.}}}

\begin{document}\cmsNoteHeader{HIN-16-020}

\hyphenation{had-ron-i-za-tion}
\hyphenation{cal-or-i-me-ter}
\hyphenation{de-vices}
\RCS$HeadURL: svn+ssh://htrauger@svn.cern.ch/reps/tdr2/notes/HIN-16-020/trunk/HIN-16-020.tex $
\RCS$Id: HIN-16-020.tex 365449 2016-08-24 18:49:38Z alverson $

\cmsNoteHeader{HIN-16-020}

\hyphenation{back-ground-sub-traction}
\newcommand{\pttrk}{\ensuremath{p_{\mathrm{T}}^{\text{trk}}}\xspace}
\newcommand{\Deltar}{\ensuremath{\Delta \mathrm{r}}\xspace}
\newcommand{\Njets}{\ensuremath{N_{\text{jets}}}\xspace}
\newcommand{\Rhodr}{\ensuremath{\mathrm{P}(\Delta\mathrm{r})}\xspace}
\newlength\cmsTabSkip\setlength{\cmsTabSkip}{1ex}

\title{Jet properties in PbPb and $\Pp\Pp$ collisions at $\sqrtsNN = 5.02\TeV$}

\date{\today}

\abstract{
Modifications of the properties of jets in PbPb collisions, relative to those in $\Pp\Pp$ collisions, are studied at a nucleon-nucleon center-of-mass energy of $\sqrtsNN = 5.02\TeV$ via correlations of charged particles with the jet axis in relative pseudorapidity ($\Delta \eta$), relative azimuth ($\Delta \phi$), and relative angular distance from the jet axis $\Deltar = \sqrt{\smash[b]{(\Delta\eta)^{2}+(\Delta\phi)^{2}}}$. This analysis uses data collected with the CMS detector at the LHC, corresponding to integrated luminosities of 404\mubinv and 27.4\pbinv for PbPb and $\Pp\Pp$ collisions, respectively. Charged particle number densities, jet fragmentation functions, and jet shapes are presented as a function of PbPb collision centrality and charged-particle track transverse momentum, providing a differential description  of jet modifications due to interactions with the quark-gluon plasma.
}

\hypersetup{
pdfauthor={CMS Collaboration},
pdftitle={Jet properties in PbPb and pp collisions at 5.02 TeV},
pdfsubject={CMS},
pdfkeywords={CMS, physics, heavy-ion collision, jets, quark gluon plasma}}

\maketitle

\section{Introduction}

Jets can be used as proxies for partons produced in the initial hard scatterings in heavy ion
collisions  to probe the properties of the quark-gluon plasma (QGP), a new state of matter
characterized by an increase in the color degrees of freedom. One of the well-established properties of the QGP is its high opacity to such penetrating probes, resulting in  significant energy loss of partons traversing the hot nuclear matter. Parton energy loss manifests itself in a variety of experimental observables, including suppression of high transverse momentum (\pt) hadrons and jets, as well as modifications of the properties of parton showers. These phenomena, collectively referred to as jet  quenching~\cite{Bjorken:1982tu}, were first observed at the BNL RHIC~\cite{Adams:2006yt, Adare:2007vu}, and subsequently at the CERN LHC~\cite{Aad:2010bu, Chatrchyan:2011sx, Chatrchyan:2012nia, Adam:2015ewa}.
The LHC experiments have previously demonstrated that the medium also affects the structure of a jet, as observed from measurements of the jet fragmentation pattern~\cite{CMS:2012vba, Aad:2014wha} and the distribution of charged-particle transverse momenta (\pttrk) as a function of the relative angular distance $\Deltar$ from the jet axis~\cite{Chatrchyan:2013kwa}, where lowercase r is used explicitly to avoid conflict with the jet clustering distance parameter $R$~\cite{fastjet}. The distance $\Deltar$ is given by $\Deltar = \sqrt{\smash[b]{(\Delta\eta)^{2}+(\Delta\phi)^{2}}}$, where $\Delta\eta$ and $\Delta\phi$ denote the relative pseudorapidity and azimuthal angle (in radians) with respect to the jet axis, respectively.  These modifications extend to large values of $\Delta\eta$ and $\Delta\phi$~\cite{mpt, HIN_2014_016, HIN_2015_011}. Various theoretical models have since attempted to account for these modifications~\cite{Apolinario:2012cg, Ayala:2015jaa, Blaizot:2014ula,Escobedo:2016jbm, Casalderrey-Solana:2016jvj}, and while most models reproduce the modification effects close to the jet axis, the large modifications far from the jet axis ($\Deltar > 0.5$) are not yet understood.

This paper describes modifications to jet structure in PbPb collisions at a nucleon-nucleon center of mass energy $\sqrtsNN = 5.02\TeV$ relative to $\Pp\Pp$ collisions at the same energy, extending previous results based on 2.76\TeV data~\cite{Chatrchyan:2013kwa,HIN_2014_016}. At the higher collision energy, an increase in the magnitude of jet quenching is expected because of the greater medium density and temperature~\cite{JetCollab:2013}, potentially increasing the size of the modification effects.  The data were collected by the CMS detector at the LHC and correspond to integrated luminosities of 404\mubinv and 27.4\pbinv for PbPb and $\Pp\Pp$ collisions, respectively.  The distributions of charged-particle tracks with respect to the jet axis are studied as a function of $\Delta\eta$, $\Delta\phi$, and $\Deltar$. The jet shapes $\rho(\Deltar)$, defined as the distribution of particle yields in $\Deltar$ weighted by \pttrk, are also examined. The results are presented differentially in \pttrk and as a function of the overlap of the colliding Pb nuclei (centrality), with head-on collisions defined as most central. Compared to the size of the data samples in Ref.~\cite{HIN_2014_016}, the present study uses a much larger data set and, hence, has a greater statistical precision. This also allows the measurements to be extended to larger distances with respect to the jet axis.

\section{The CMS detector}
\label {sec:Detector}

The central feature of the CMS apparatus is a superconducting solenoid of 6\unit{m} internal diameter, providing a magnetic field of 3.8\unit{T}. Within the solenoid volume are a silicon pixel and strip tracker, a lead tungstate crystal electromagnetic calorimeter (ECAL), and a brass and scintillator hadron calorimeter (HCAL), each composed of barrel and endcap sections.  Two hadronic forward (HF) steel and quartz-fiber calorimeters complement the barrel and endcap detectors, extending the calorimeter from the range $\abs{\eta}<3.0$ provided by the barrel and endcap out to $\abs{\eta}<5.2$. The scalar \pt sum of calorimeter towers in the HF region~(\mbox{$4.0<\abs{\eta}<5.2$}) is used to define the event centrality in PbPb events and to divide the event sample into centrality classes, each representing a percentage of the total nucleus-nucleus hadronic interaction cross section.  A detailed description of the centrality determination can be found in Ref.~\cite{Chatrchyan:2012nia}.

Jets are reconstructed within the range $\abs{\eta}<1.6$. In the region $\abs{\eta}<1.74$, the HCAL cells have widths of 0.087 in both $\eta$ and $\phi$ and thus provide high granularity. Within the central barrel region of $\abs{\eta}<1.48$, the HCAL cells map onto $5\times5$ ECAL crystal arrays to form calorimeter towers projecting radially outwards from the nominal interaction point. Within each tower, the energy deposits in ECAL and HCAL cells are summed to define the calorimeter tower energies, which are subsequently clustered to reconstruct the jet energies and directions~\cite{Chatrchyan:2011ds}.

The CMS silicon tracker measures charged-particle tracks within $\abs{\eta}<2.5$. It consists of 1440 silicon pixel and 15\,148 silicon strip detector modules. For charged particles with $1<\pt<10\GeV$ in the barrel region, the track resolutions are typically 1.5\% in \pt and 25--90 (45--150)\mum in the impact parameter direction transverse (longitudinal) to the colliding beams~\cite{TRK-11-001}.
A detailed description of the CMS detector, together with a definition of the coordinate system used and the relevant kinematic variables, can be found in Ref.~\cite{Chatrchyan:2008zzk}.

\section{Event selection and simulated event samples}

The $\Pp\Pp$ and PbPb data are selected with a calorimeter-based trigger that uses the anti-\kt jet clustering algorithm with distance parameter of $R=0.4$~\cite{bib_antikt}. The trigger requires events to contain at least one jet with $\pt>80\GeV$.  This trigger is fully efficient for events containing jets with reconstructed $\pt>90\GeV$. For both PbPb and $\Pp\Pp$ collisions, the data selected by this trigger are referred to as "jet-triggered", and are used to study the jet-related particle yields. While jet-triggered samples are used in $\Pp\Pp$ collisions to correct for the limited jet and track acceptance via an event mixing technique described in Section~\ref{sec:me_corr}, an additional data sample collected with a minimum-bias trigger~\cite{CMS:mb} is used for this correction in PbPb collisions in order to properly capture the long-range correlated particle yields in PbPb data.  To reduce contamination from noncollision events, including calorimeter noise and beam-gas collisions, vertex and noise filters are applied to both the $\Pp\Pp$ and PbPb data as described in previous analyses~\cite{Chatrchyan:2011sx,Chatrchyan:2012nia}.  The filters include the requirements that events  contain at least 3\GeV of energy in each of the two HF calorimeters and that a primary vertex with at
  least two tracks be present within 15\unit{cm} of the center of the nominal interaction region along the beam axis ($\abs{v_{z}}<15\unit{cm}$).

Monte Carlo (MC) simulated event samples are used to evaluate the performance of the event reconstruction, particularly the track reconstruction efficiency and the jet energy response and resolution. The MC samples use the \PYTHIA (version 6.424, tune Z2~\cite{bib_pythia,Field}) event generator to describe the hard scattering, parton showering, and hadronization of the partons.
The \GEANTfour~\cite{bib_geant} toolkit is used to simulate the CMS detector response. To account for the soft underlying PbPb event component, the hard \PYTHIA interactions are embedded into simulated minimum-bias PbPb events, produced with the \HYDJET 1.383~\cite{Lokhtin:2005px} event generator. We refer to this latter sample as \PYTHIA+\HYDJET.

In PbPb collisions, jets are produced  more frequently in central events than in noncentral events because of the large number of binary collisions per nuclear interaction. Since the \HYDJET event generator simulates minimum-bias PbPb collisions only, a centrality-based reweighting is applied to the \PYTHIA+\HYDJET sample in order to match the centrality distribution of the jet-triggered PbPb data. An additional reweighting procedure is performed to match the simulated $v_{z}$ distributions to data for both the $\Pp\Pp$ and PbPb samples.

\section{Jet and track reconstruction}
\label{jet_reco}

The jet reconstruction in PbPb and $\Pp\Pp$ events is performed with the anti-\kt jet algorithm with a distance parameter of $R=0.4$, as implemented in the \FASTJET framework~\cite{fastjet}, with individually calibrated calorimeter towers as input. Only calorimeter information is used in order to avoid biases due to an interplay between track reconstruction efficiency and the jet energy scale. In PbPb collisions, the contributions of the underlying event are subtracted using a variant of the iterative "noise/pedestal subtraction" technique described in Ref.~\cite{Kodolova:2007hd}.  Following the subtraction, jets are calibrated such that the calorimeter response is uniform as a function of jet \pt and $\eta$. A reweighting procedure based on the number of combined calorimeter towers and associated charged-particle tracks with $\pt>2\GeV$ is then applied to correct for the variation in detector response with the total number of jet constituents.
This latter calibration corrects for a difference in the simulated calorimetric jet energy response between quark and gluon jets and reduces the difference in this response between the two jet flavors, as a function of jet \pt, from 10\% to around 3\%.
After reconstruction and offline jet energy calibration, jets are required to have $\pt>120\GeV$ and $\abs{\eta}<1.6$. Within this selection, it is possible for multiple jets to be selected from the same event. Roughly 25\% of $\Pp\Pp$ events contain multiple jets that satisfy all kinematic selection criteria and no additional selections are made to distinguish between leading and subleading jets in this measurement.

For $\Pp\Pp$ data and simulations, charged-particle tracks are reconstructed using an iterative tracking method~\cite{TRK-11-001} that allows the reconstruction of charged-particle tracks within $\abs{\eta}<2.4$  down to $\pt=0.1\GeV$. For the PbPb case, an alternative iterative charged-particle reconstruction procedure is employed because of the large track multiplicity in such collisions, as discussed in earlier heavy ion analyses~\cite{CMS:2012aa,Chatrchyan:2013kwa}. This reconstruction algorithm is based on hit information from both the pixel and silicon strip subdetectors and is capable of reconstructing charged-particle tracks down  to $\pt=0.4\GeV$. The tracking efficiency in $\Pp\Pp$ collisions ranges from approximately 80\% at $\pt=0.5\GeV$ to 90\% or higher for $\pt>10\GeV$. Track reconstruction is more difficult in the heavy ion environment because of the large track multiplicity and so the tracking efficiency ranges from approximately 30\% at $\pt=0.5\GeV$ to about 70\% at $\pt=10\GeV$~\cite{HIN-15-015}.  Corrections for the tracking efficiency and related effects are derived as a function of \pttrk, $\eta$, and $\phi$ using \PYTHIA simulations, and, additionally for PbPb events, as a function of centrality using \PYTHIA+\HYDJET.

\section{Jet--track angular correlations}
\label{sec:me_corr}

Correlations between reconstructed jets and charged-particle tracks are studied by forming a two-dimensional array of the $\Delta\eta$ and $\Delta\phi$ values of the tracks relative to the jet axis. Events in PbPb collisions are divided into four centrality intervals based on the fraction of the total recorded energy that is collected within the HF calorimeter, given by 0--10\% (most central, corresponding to the largest overlap of the colliding nuclei), 10--30\%, 30--50\%, and 50--100\% (most peripheral), and into eight bins of \pttrk bounded by the values 0.7, 1, 2, 3, 4, 8, 16, 20, and 300\GeV. After classification, the jet--track correlation yields are normalized by the number of jets in the sample and corrected for tracking efficiency on a per-track basis. The normalization creates a per-jet averaged $\Delta\eta$--$\Delta\phi$ distribution of charged particles about the jet axis for each \pttrk and centrality interval.  For the \pttrk measurements, the two-dimensional correlations are weighted by \pttrk on a per-track basis, producing a per-jet averaged $\Delta\eta$--$\Delta\phi$ distribution of \pttrk  about the jet axis.

Following construction of the  two-dimensional correlations described above, the remaining analysis procedure consists of the following three steps: first, an acceptance correction is derived using a mixed-event method, which accounts for the effects of the limited detector acceptance.  Second, the backgrounds from tracks unrelated to jets are subtracted using a region in $\Delta\eta$ far from the jet axis. Third, simulation-derived corrections are applied to account for jet reconstruction biases. Each of these steps is described in detail below.

As a consequence of the shape of the jet and track $\eta$ distributions and the different acceptance requirements for jets ($\abs{\eta}<1.6$) and tracks ($\abs{\eta}<2.4$), the jet--track correlation yield decreases with increasing $\Delta\eta$. This decrease is due to a combination of the $\eta$-dependence of jet and particle production, as well as the limited detector acceptance. Jet--track pairs, therefore, have a higher probability to be reconstructed at central values of jet and/or track $\eta$ than in the forward or backward regions.
To correct for this pair-acceptance effect, a mixed-event distribution is constructed by creating jet--track pairs using jets from the jet-triggered event sample and tracks from a sample of minimum-bias events, matched in the vertex position along the beam axis (within 1\unit{cm}) and collision centrality (within 2.5\%). This procedure is analogous to that employed for two-particle correlations in Refs.~\cite{ppRidge,HIN-11-001,pPbridge} and for jet--track correlations in Refs.~\cite{HIN13002,HIN_2014_016, HIN_2015_011}.  In the following, \Njets denotes the total number of jets satisfying the selection criteria, either from $\Pp\Pp$ or PbPb collisions. The per-jet associated yield is corrected for the acceptance effects via the following relation:
  \begin{equation}
  \label{eq:2pcorr_incl}
  \frac{1}{\Njets}\frac{\rd^{2} N}{\rd\Delta\eta\, \rd\Delta\phi}
  = \frac{ME(0,0)}{ME(\Delta\eta,\Delta\phi)}\ S(\Delta\eta,\Delta\phi),
  \end{equation}
  where the signal pair distribution, $S(\Delta\eta,\Delta\phi)$,
  represents the yield of jet--track pairs from the same event, normalized by $\Njets$,
  \begin{equation}
  \label{eq:signal}
  S(\Delta\eta,\Delta\phi) = \frac{1}{\Njets}\frac{\rd^{2} N^{\text{same}}}{\rd\Delta\eta\, \rd\Delta\phi}\ ,
  \end{equation}
 and the mixed-event pair distribution $ME(\Delta\eta,\Delta\phi)$ is
  \begin{equation}
  \label{eq:background}
  ME(\Delta\eta,\Delta\phi) = \frac{1}{\Njets}\frac{\rd^{2}N^{\text{mix}}}{\rd\Delta\eta\, \rd\Delta\phi}\ .
  \end{equation}
The ratio $ME(0,0)/ME(\Delta\eta,\Delta\phi)$ is the normalized correction factor;  $ME(0,0)$  is the mixed-event yield for jet--track pairs that are approximately collinear. These collinear  pairs have  the maximum  pair acceptance.

The acceptance-corrected distributions resulting from Eq.~(\ref{eq:2pcorr_incl}) exhibit a Gaussian-like peak confined to a fairly narrow $\Delta\eta$, $\Delta\phi$ range on top of a large background from unrelated jet--track pairs and pairs connected through long-range correlations, \eg, azimuthal anisotropies.  To model this background, the $\Delta\phi$ distribution averaged over $1.5<\abs{\Delta\eta}<2.5$ is used to estimate the $\Delta\phi$-dependence of the combinatoric contribution to the correlations over the entire  $\abs{\Delta\eta}<4.0$ region and is subtracted from the acceptance-corrected yields of Eq.~(\ref{eq:2pcorr_incl}).
The use of a narrow range in $\Delta\eta$ to model the background automatically accounts for the magnitude of the azimuthal anisotropy contribution, without the need for an explicit measurement.

Finally, simulation-based corrections are applied to account for jet position resolution and two biases in the jet reconstruction: a bias toward selecting jets with a harder constituent \pt spectrum (affecting PbPb and $\Pp\Pp$ events similarly), and a bias toward selecting jets that are affected by upward fluctuations in the \pt values of the objects in the soft underlying event (relevant for PbPb events only). Jets with a harder constituent \pt spectrum are more likely to be successfully reconstructed than jets with a softer constituent \pt spectrum because the calorimeter response does not scale linearly with incident particle energy, resulting in a bias toward the selection of jets with fewer associated tracks.  This bias is reduced, but not eliminated, by applying the jet energy  corrections mentioned above based on the number of jet constituents.  A residual correction for this bias is derived following the method described in Refs.~\cite{mpt,HIN_2014_016,HIN_2015_011}, by comparing per-jet yields of generated particles correlated to reconstructed jets relative to those correlated to generated jets. For $\Pp\Pp$ events, this correction is derived using the \PYTHIA simulation. For PbPb events, we consider only generated particles from the embedded \PYTHIA hard process, excluding particles from the underlying event.

For PbPb events, there is an additional jet reconstruction bias toward the selection of jets that are produced in the vicinity of upward fluctuations in the underlying event. Since the jet \pt spectrum falls steeply, more jets on upward fluctuations are included in the sample than jets on downward fluctuations are excluded.  To estimate and then account for this bias, we follow a similar procedure to that outlined in Refs.~\cite{Chatrchyan:2012gw, HIN_2014_016, HIN_2015_011}.  We consider correlations in the \PYTHIA+\HYDJET sample between reconstructed jets and generated particles from only the \HYDJET underlying event, excluding particles from the embedded hard process. To avoid propagating low-\pt \HYDJET fluctuations to the data, the per-jet excess yields from the \HYDJET underlying event are fit with Gaussian functions in both $\Delta\eta$ and $\Delta\phi$ and applied as a correction to the PbPb data.

\section{Systematic uncertainties}
\label{sec:systematics}

A number of sources of systematic uncertainty are considered, including effects from the tracking efficiency, acceptance corrections, background subtraction, and jet reconstruction.  Where relevant, systematic uncertainties related to tracking and jet reconstruction are determined by comparing properties of reconstructed \PYTHIA and \PYTHIA+\HYDJET events to their generated counterparts. Jet reconstruction related sources of systematic uncertainty include the two biases in jet reconstruction discussed in Section~\ref{sec:me_corr} and the jet energy scale (JES).

To evaluate the uncertainty related to the jet reconstruction biases and the jet position resolution, three variations of the analysis are performed to gauge differences of the jet energy calorimetric response in data relative to the MC simulation. First, the collision reaction plane dependence of the jet reconstruction performance is tested by determining the corrections independently for in-plane and out-of-plane jets, where the reaction plane is the plane containing   the beam axis and the projection onto the transverse   plane of the line connecting the centers   of the colliding nuclei.  The difference in the corrections and hence the resulting uncertainty is found to be negligible.  Second, the uncertainties associated with the evaluation of the magnitude of both jet reconstruction biases are obtained by varying the corresponding acceptance correction and background subtraction within their statistical uncertainties. For the jet constituent hardness bias, the resulting uncertainty in the correction factor is found to be negligible. Conversely, for the correction accounting for upward underlying event fluctuations, the uncertainty is found to vary between 1 and 18\% of the total correction, strongly depending on \pttrk and the centrality. Third, the dependence on the relative numbers of quark and gluon jets is studied. By fitting distributions of the quark and gluon jet constituent multiplicities in simulation, we obtain templates that are used in a fit to estimate the fraction of quark and gluon jets in data.  We observe that the quark jet fraction in data is centrality dependent, ranging between 49 and 56\%, while the quark jet fraction in simulation is constant at approximately  42\%. The jet reconstruction bias corrections are reweighted in simulation to correspond to the measured ratio of quark to gluon jets. The reweighting has a negligible impact on the correction accounting for upward fluctuations in the underlying event, but affects the corrections that account for the bias toward jets with a harder constituent \pt spectrum on the order of 10\%, resulting in an uncertainty of 2\% in the correlated yields.
In addition to the uncertainties evaluated from these three variations of the analysis, an uncertainty associated with the Gaussian fitting procedure described at the end of Section~\ref{sec:me_corr} is given by the uncertainty in this fit, though this uncertainty is negligible compared to the overall reconstruction bias uncertainty.

The JES is subject to three sources of uncertainty, all of roughly equal magnitude. All are related to potential differences between simulation and data. First, differences in the relative fraction of quark and gluon jets can affect the overall energy scale of the jet sample. Second, jet quenching effects in PbPb events are not simulated in \HYDJET. Third, residual differences in the calorimeter response exist between data and simulation.
All three sources are conservatively accounted for by a 5\% variation in the jet energy scale. Template fits are used to estimate the quark and gluon jet fractions in $\Pp\Pp$ data, absent of quenching effects, and are found to be consistent to within 5\%. Studies of jet quenching result in an estimate of a 7--10\GeV shift in jet energy, roughly half of which remains in the jet cone. Finally, residual differences in calorimeter response are estimated using gamma-jet correlations~\cite{mpt}. These studies yield observed discrepancies of up to 5\% in the calorimeter response between photons and jets.  To evaluate the effect of all these uncertainties on the results, we vary the jet \pt threshold of 120\GeV up and down by 6\GeV, corresponding to a shift of 5\%. The resulting uncertainty in the correlated yields is found to be around 2\%, because the jet multiplicities vary only slowly as a function of jet \pt.

The uncertainty related to the tracking efficiency is estimated from simulation by taking the ratio between the corrected reconstructed yields and the corresponding
  generated yields. This ratio is referred to as the ``closure" and its deviation from unity defines the systematic uncertainty.  The systematic uncertainty
  in the tracking  correction is found to be 1\% in both PbPb and $\Pp\Pp$ events, noting that the PbPb closure is derived using simple two-dimensional efficiency tables in $\eta$ and $\phi$, while the $\Pp\Pp$ closure uses a multidimensional parameterization of the tracking efficiency. An uncertainty is also evaluated to account for possible differences in track reconstruction between data and simulation, including the erroneous reconstruction of tracks, and is found to be 5\% (4\%) in PbPb ($\Pp\Pp$) events.

The uncertainty arising from long-range $\Delta\eta$-dependent asymmetries of the mixed-event acceptance correction is estimated by considering the sideband asymmetry of the correlated yield after dividing by the mixed-event background.  Ideally, the mixed-event acceptance corrected signal is composed of a correlated jet--track yield sitting on top of an uncorrelated background. Far from the jet axis, the uncorrelated background should dominate the mixed event corrected signal and, thus, the signal is expected to be uniform for $\abs{\Delta\eta}>1.5$. To probe for discrepancies from this ideal case, each sideband region of the final $\Delta\eta$ distribution ($-2.5<\Delta\eta<-1.5$ and $1.5<\Delta\eta<2.5$) is independently fit with a constant to estimate the nonuniformity of the mixed event correction, with respect to any residual $\Delta\eta$ asymmetry.  The difference between these two fits is assigned as a systematic uncertainty, and is found to be about 5\% for the lowest \pttrk bin, where such effects are largest.

Uncertainties associated with the background subtraction are evaluated by considering the average point-to-point difference between two sideband regions ($1.5<\abs{\Delta\eta}<2.0$ and $2.0<\abs{\Delta\eta}<2.5$) following the background subtraction. In central events (0--10\%), the background subtraction uncertainty is found to be roughly 4\% for the lowest \pttrk bin, where the ratio of signal to background events is lowest, and decreases as a function of centrality and \pttrk.

The systematic uncertainties from all sources are added in quadrature. Table~\ref{tab:sys} lists the ranges of the estimated contributions from the individual sources.

\begin{table}[htbp]
\centering
\topcaption{Systematic uncertainties in percentage for the measurements of the jet--track correlations in PbPb and $\Pp\Pp$ events. The PbPb results are given in intervals of centrality. Where an uncertainty range is given, the upper edge of the range corresponds to the bin with the smallest \pttrk values.}
\begin{tabular}{cccccc}
\hline
& \multicolumn{4}{c}{PbPb centrality intervals} & $\Pp\Pp$ \\
Source                                   & 0--10\% &    10--30\% &   30--50\%&    50--100\% &  \\
\hline
Background fluctuation bias              & 1--15    & 0--11 &   0--5 &    0--2 &    \NA \\
Jet constituent \pt bias                 & 2        & 2 &        2 &         2    &    2 \\
Residual JES                             & 4 &        4 &        4 &        4 &       4 \\
Tracking efficiency                      & 1 &        1 &         1 &         1 &       1  \\
Residual tracking efficiency             & 5 &        5 &         5 &         5 &       4 \\
Pair-acceptance corrections              & 1--4 &     1--2 &     1--2 &     1--2 &    1--2 \\
Background subtraction                   & 0--4 &     0--4 &     0--4 &     0--3 &     0--3 \\[\cmsTabSkip]
Total                                    & 7--17    & 7--14     & 7--10   & 7--8 &    6--7 \\
\hline
\end{tabular}
\label{tab:sys}
\end{table}

\section{Results}

\begin{figure}[hb!]
     \centering
	\includegraphics[width=0.99\textwidth]{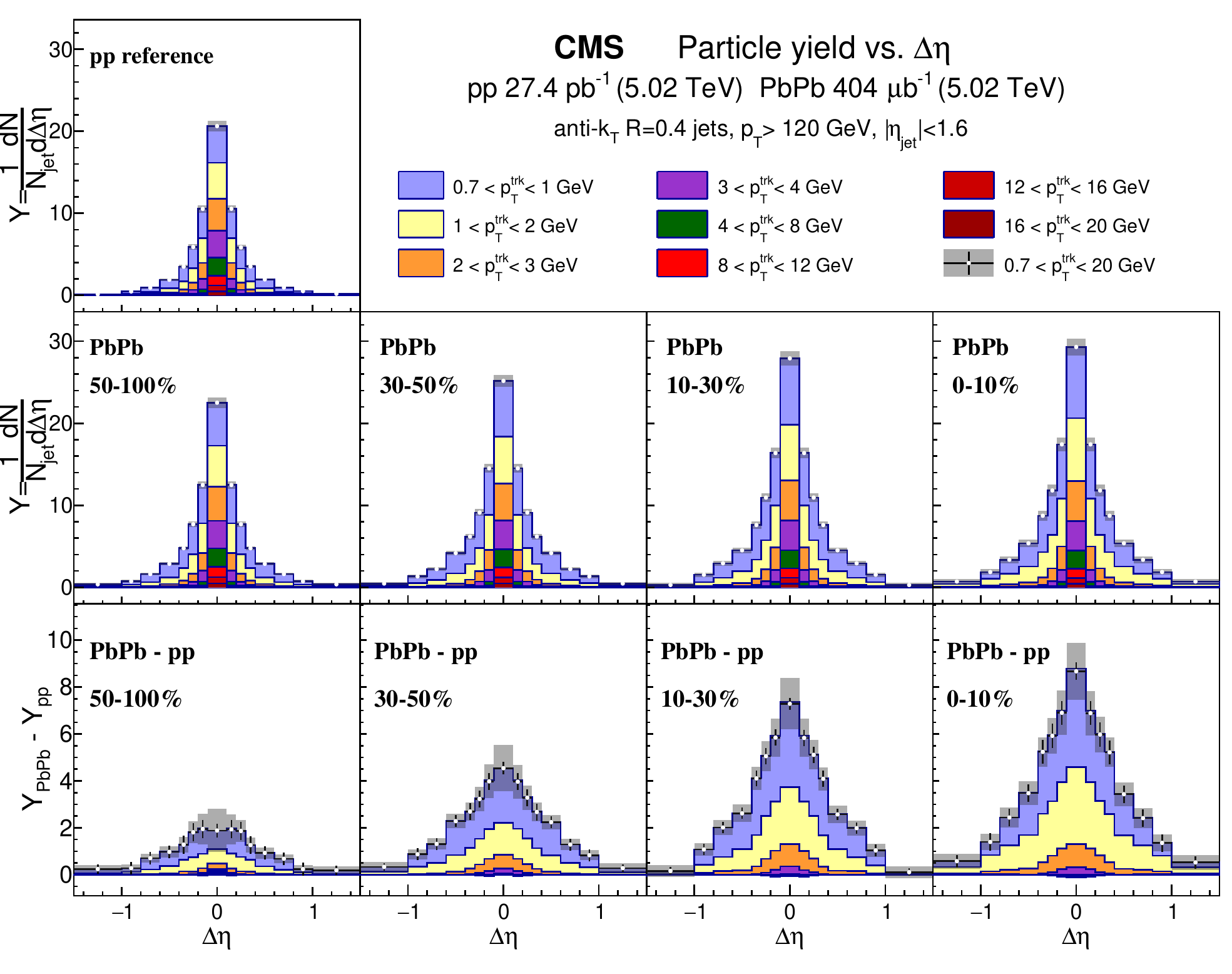}
         \caption{The distribution of jet-correlated charged-particle tracks with $\abs{\Delta\phi}<1.0$ as a function of $\Delta\eta$ in $\Pp\Pp$ (top left) and PbPb (middle row) collisions. The PbPb results are shown for different centrality   regions. The bottom row shows the difference between the PbPb and $\Pp\Pp$ data. The intervals in \pttrk are indicated by the stacked histograms. The inclusive points ($0.7<\pttrk<20$\GeV) are shown by open white circles. The grey bands surrounding these points show the total systematic uncertainties.}
       \label{fig:yield_deta_stacked}
 \end{figure}

\begin{figure}[hb!]
    \centering
    \includegraphics[width=0.99\textwidth]{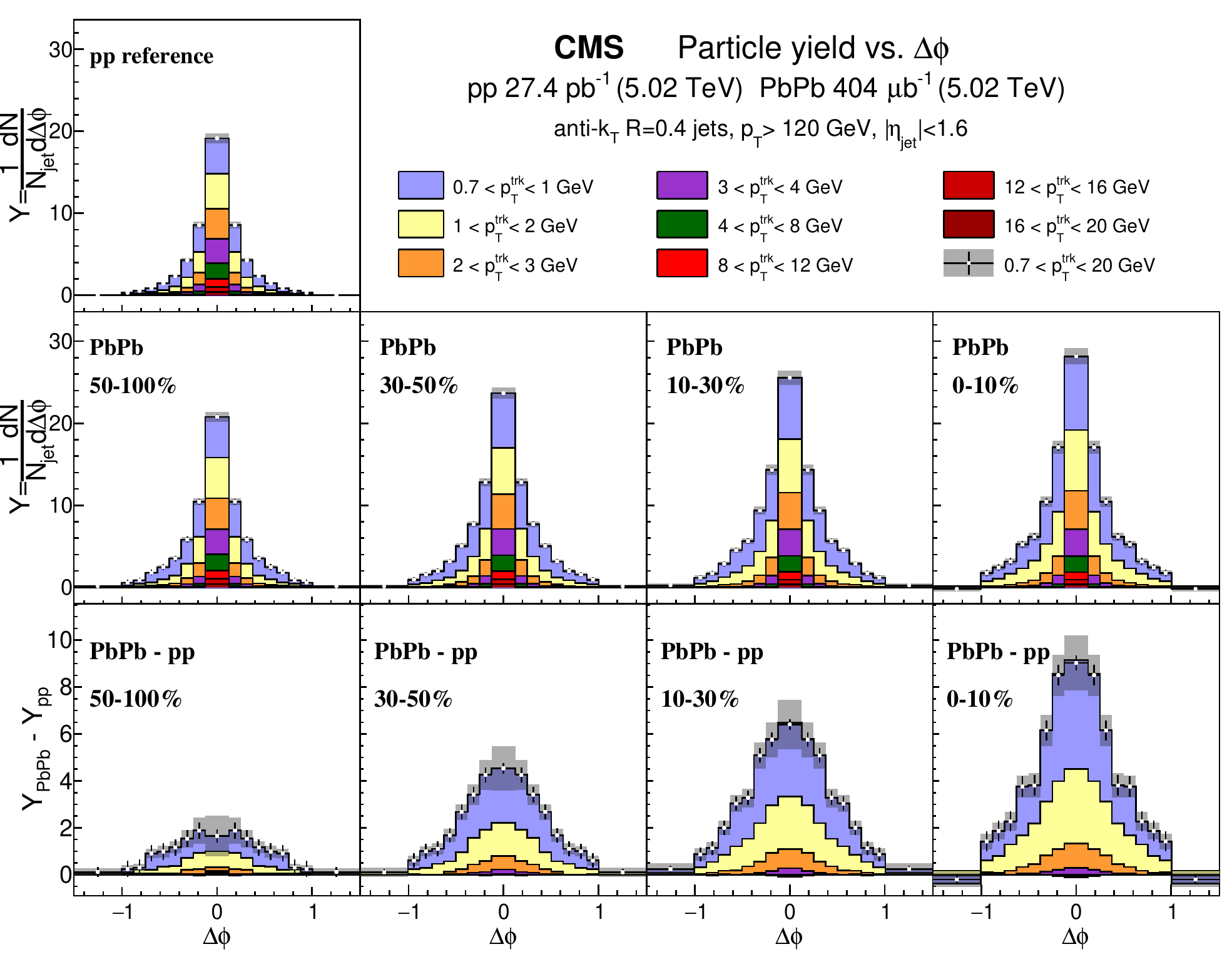}
         \caption{The distribution of jet-correlated charged-particle tracks with $\abs{\Delta\eta}<1.0$ as a function of $\Delta\phi$ in $\Pp\Pp$ (top left) and PbPb (middle row) collisions. The PbPb results are shown for different centrality   regions. The bottom row shows the difference between the PbPb and $\Pp\Pp$ data. The intervals in \pttrk are indicated by the stacked histograms. The inclusive points ($0.7<\pttrk<20$\GeV) are shown by open white circles. The grey bands surrounding these points show the total systematic uncertainties.}
       \label{fig:yield_dphi_stacked}
 \end{figure}

Figures~\ref{fig:yield_deta_stacked} and~\ref{fig:yield_dphi_stacked} present the PbPb and $\Pp\Pp$ charged-particle track yields as a function of $\Delta\eta$ and $\Delta\phi$, respectively. The results are shown for the different centrality regions, with stacked histograms to indicate the intervals in \pttrk, and open circles to denote the inclusive summed yields. To illustrate the dependence of the results on the medium, the difference between the PbPb and $\Pp\Pp$ data are shown in the bottom row of each figure. Note the inclusive points shown by the open circles on this row are the sum of both the positive and negative contributions. The distributions exhibit an enhancement of soft particles ($\pttrk<3$\GeV) in the PbPb data, relative to the $\Pp\Pp$ data, that becomes more pronounced with increasing centrality.  This low-\pttrk excess is relatively symmetric in $\Delta\eta$ and $\Delta\phi$ and remains significant even for large $\Delta\eta$ and $\Delta\phi$.  The particle yield excess in the PbPb data decreases with increasing \pttrk, such that no significant enhancement is observed for $\pttrk>3\GeV$.

\begin{figure}[ht!]
       \centering
	\includegraphics[width=0.99\textwidth]{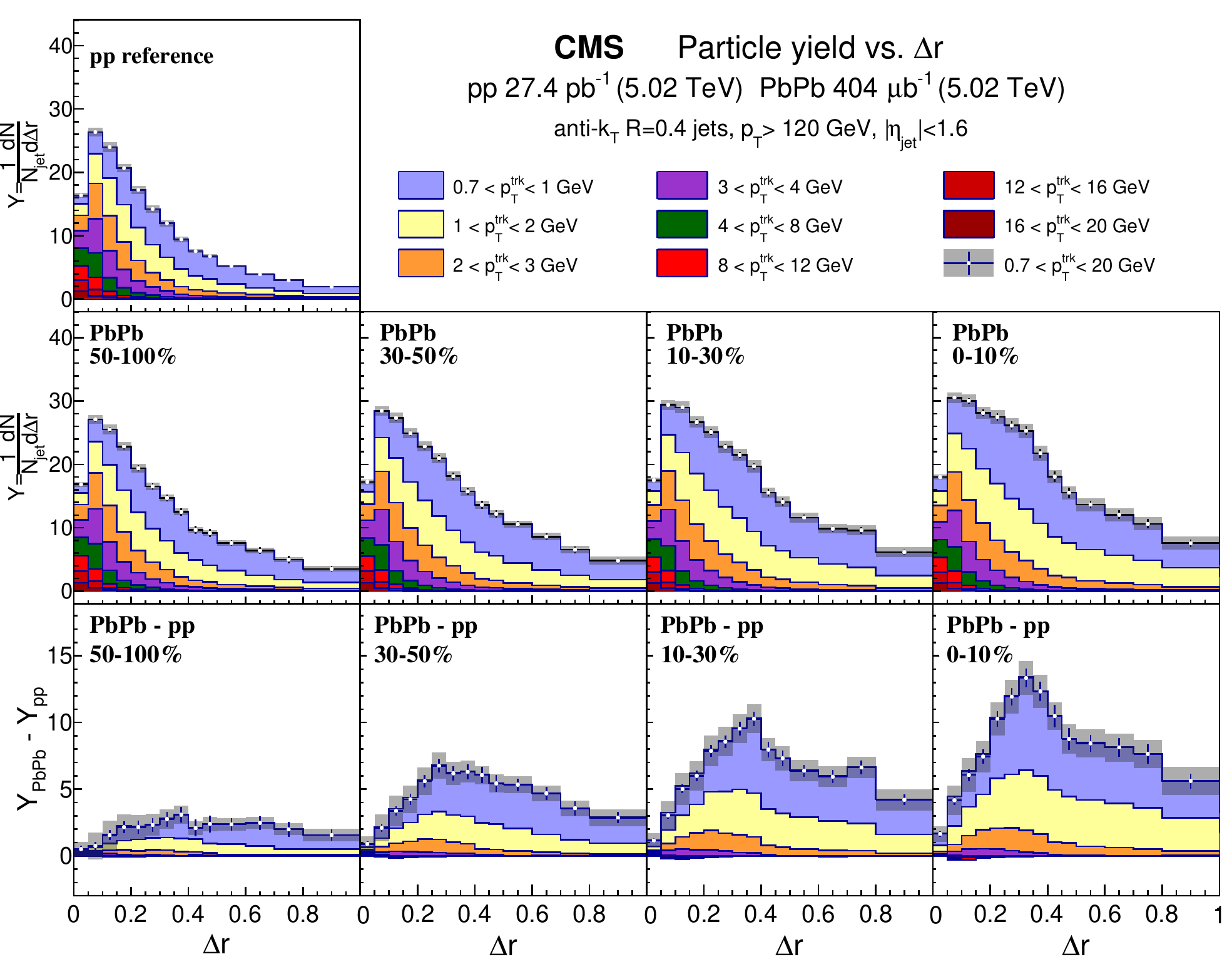}
         \caption{The distribution of jet-correlated charged-particle tracks as a function of $\Deltar$ in $\Pp\Pp$ (top left) and PbPb (middle row) collisions. The PbPb results are shown for different centrality   regions. The bottom row shows the difference between the PbPb and $\Pp\Pp$ data. The intervals in \pttrk are indicated by the stacked histograms. The inclusive points ($0.7<\pttrk<20$\GeV) are shown by open white circles. The grey bands surrounding these points show the total systematic uncertainties.}
       \label{fig:yield_dr_stacked}
 \end{figure}

Figure~\ref{fig:yield_dr_stacked} shows the jet--track correlations as a function of $\Deltar$. As for the $\Delta\eta$ and $\Delta\phi$ distributions, an enhancement for $\pttrk>3\GeV$ is seen in the PbPb data relative to the $\Pp\Pp$ data, which increases with increasing centrality. The change of slope in the figures at $\Deltar = 0.4$ is due to the nature of the anti-\kt algorithm, where low-\pt particles just inside the jet radius parameter are generally clustered within the jet and particles just outside the radius are not.

\begin{figure}[ht!]
    \centering
       \includegraphics[width=0.99\textwidth]{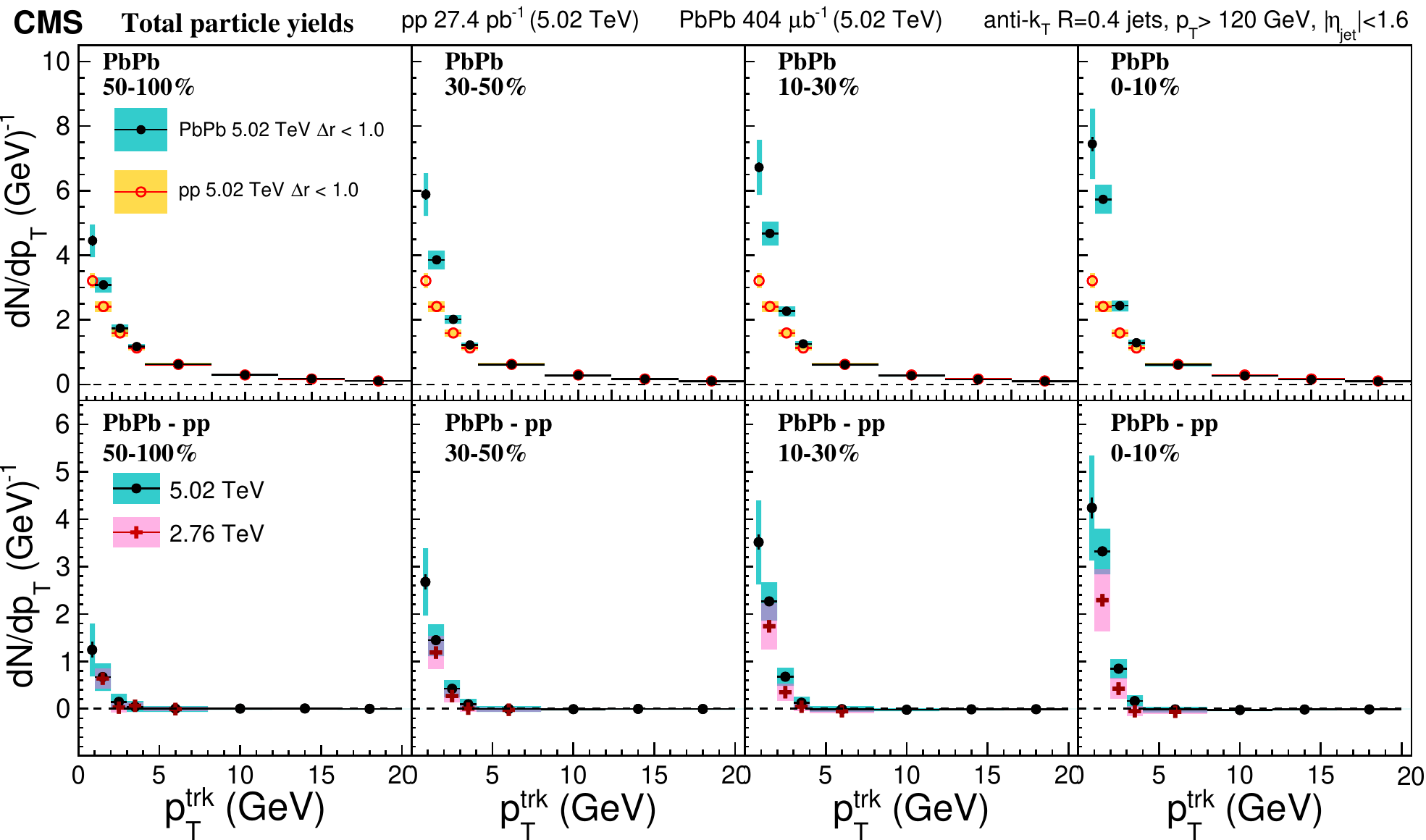}
         \caption{The distribution of jet-correlated charged-particle tracks with $\Deltar <1$ (top row) as a function of \pttrk in PbPb and $\Pp\Pp$ collisions. The PbPb results are shown for different centrality regions. Statistical uncertainties are shown by the vertical error bars and systematic uncertainties by the shaded boxes.           The corresponding results for the difference between  the PbPb and $\Pp\Pp$ data  are shown in the bottom row, with the analogous   difference observed at 2.76\TeV~\cite{HIN_2014_016}.}
       \label{fig:yield_integral}
 \end{figure}

To further illustrate the \pttrk dependence of the results, Fig.~\ref{fig:yield_integral} (top row) shows the charged-particle track yields in PbPb and $\Pp\Pp$ events as a function of \pttrk, integrated over $\Delta\eta$ and $\Delta\phi$. The excess observed in PbPb events is seen to decrease smoothly with increasing \pttrk, in each centrality interval.
For $\pttrk>3\GeV$, the PbPb results are consistent with those from the $\Pp\Pp$ collisions.  Figure~\ref{fig:yield_integral} also shows a comparison of the results at 5.02\TeV with those previously obtained at 2.76\TeV~\cite{HIN_2014_016} in order to show the dependence of the excess on the center-of-mass energy. The results from the two center-of-mass energies are consistent within one standard deviation.

\begin{figure}[hb!]
    \centering
       \includegraphics[width=0.99\textwidth]{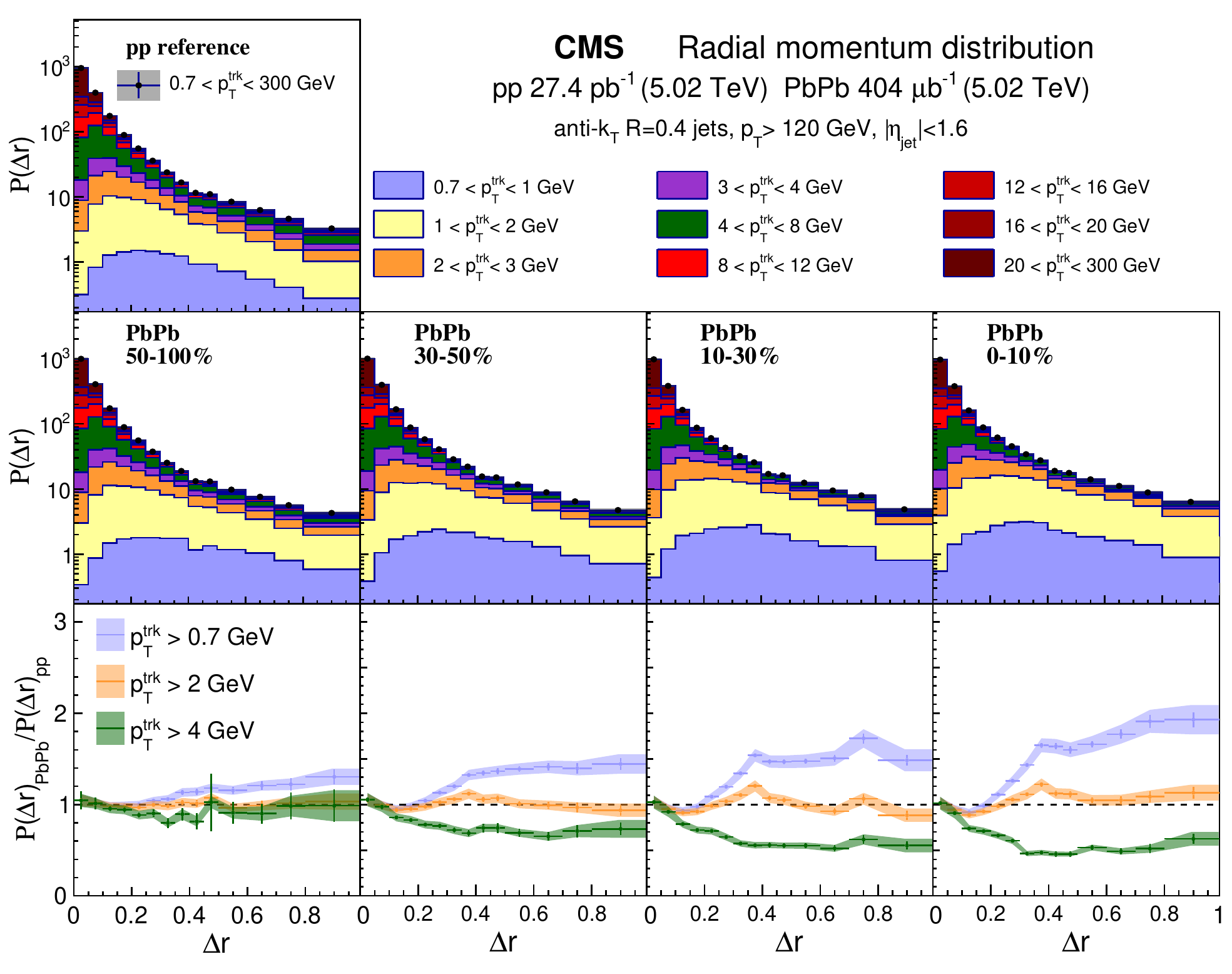}
         \caption{The radial jet momentum distribution $\Rhodr$ of jets in $\Pp\Pp$ (top left) and PbPb (middle row) collisions. The PbPb results are shown for different centrality regions. The bottom row shows the ratio between PbPb and $\Pp\Pp$ data for the indicated intervals of \pttrk. The shaded bands show the total systematic uncertainties.}
       \label{fig:profile_dr_stacked}
 \end{figure}

Measurements of the jet shapes $\rho(\Deltar)$ are obtained by examining the distribution of charged-particle tracks in annular rings of width $\Deltar = 0.05$ around the jet axis, with each particle weighted by its \pttrk value. In contrast to Figs.~\ref{fig:yield_deta_stacked}-\ref{fig:yield_integral}, tracks with $\pttrk>20\GeV$ are included in these distributions as they make a significant contribution to the jet momentum, even though their rate is small. The resulting transverse momentum profile \Rhodr of the jet is defined as:
\begin{equation}
\Rhodr = \frac{1}{\delta \mathrm{r}} \frac{1}{N_{\rm jets}} \large{\Sigma_{\text{jets}}} \Sigma_{\text{tracks} \in ( \Delta \mathrm{r_{a}}, \Delta \mathrm{r_{b}} )} \pttrk, \, \Deltar<1,
\end{equation}
where $\Delta \mathrm{r_{a}}$ and $\Delta \mathrm{r_{b}}$ define the annular edges of $\Deltar$, and $\delta \mathrm{r}$ = $\Delta \mathrm{r_{b}} - \Delta \mathrm{r_{a}}$. This profile is normalized to unity within $\Deltar = 1$ to produce $\rho(\Deltar)$:
\begin{equation}
\rho(\Deltar) = \frac{\mathrm{P} (\Deltar)}{\large{\Sigma_{\text{jets}}} \Sigma_{\text{tracks}} \pttrk}.
\end{equation}
The plots in the top left and middle row of Fig.~\ref{fig:profile_dr_stacked} show the $\Rhodr$ distribution for $\Pp\Pp$ and PbPb events, respectively. The bottom row shows the ratio of the PbPb to the $\Pp\Pp$ data for three different ranges in \pttrk, namely $\pttrk>0.7$, 2.0, and 4.0\GeV.
These results demonstrate a large excess of  soft particles in PbPb events relative to $\Pp\Pp$ events  at intermediate to large angles from the jet axis,  compensated for by a relative depletion at all angles  of tracks at high-\pttrk. For charged-particle tracks with $\pttrk>0.7$\GeV, the difference between the PbPb and $\Pp\Pp$ data reaches nearly a factor of two for large $\Deltar$ in central collisions, as seen from the bottom right plot in Fig.~\ref{fig:profile_dr_stacked}. This behavior is possibly due to a combination of jet quenching in the medium and the so called "backreaction" response of the medium to the jet, where the jet induces spallation-type effects in the soft underlying event, increasing low-\pttrk contributions over a wide range in $\Deltar$. Out of a number of theoretical models at 2.76\TeV~\cite{taichibana:2017, Vitev, Chien:2015, Casalderrey-Solana:2016jvj} only those that include such effects in combination with other processes, like jet quenching and jet broadening, are able to reproduce the large low-\pttrk excesses at very large $\Deltar$ observed in the momentum profiles.

\begin{figure}[hb!]
    \centering
       \includegraphics[width=0.99\textwidth]{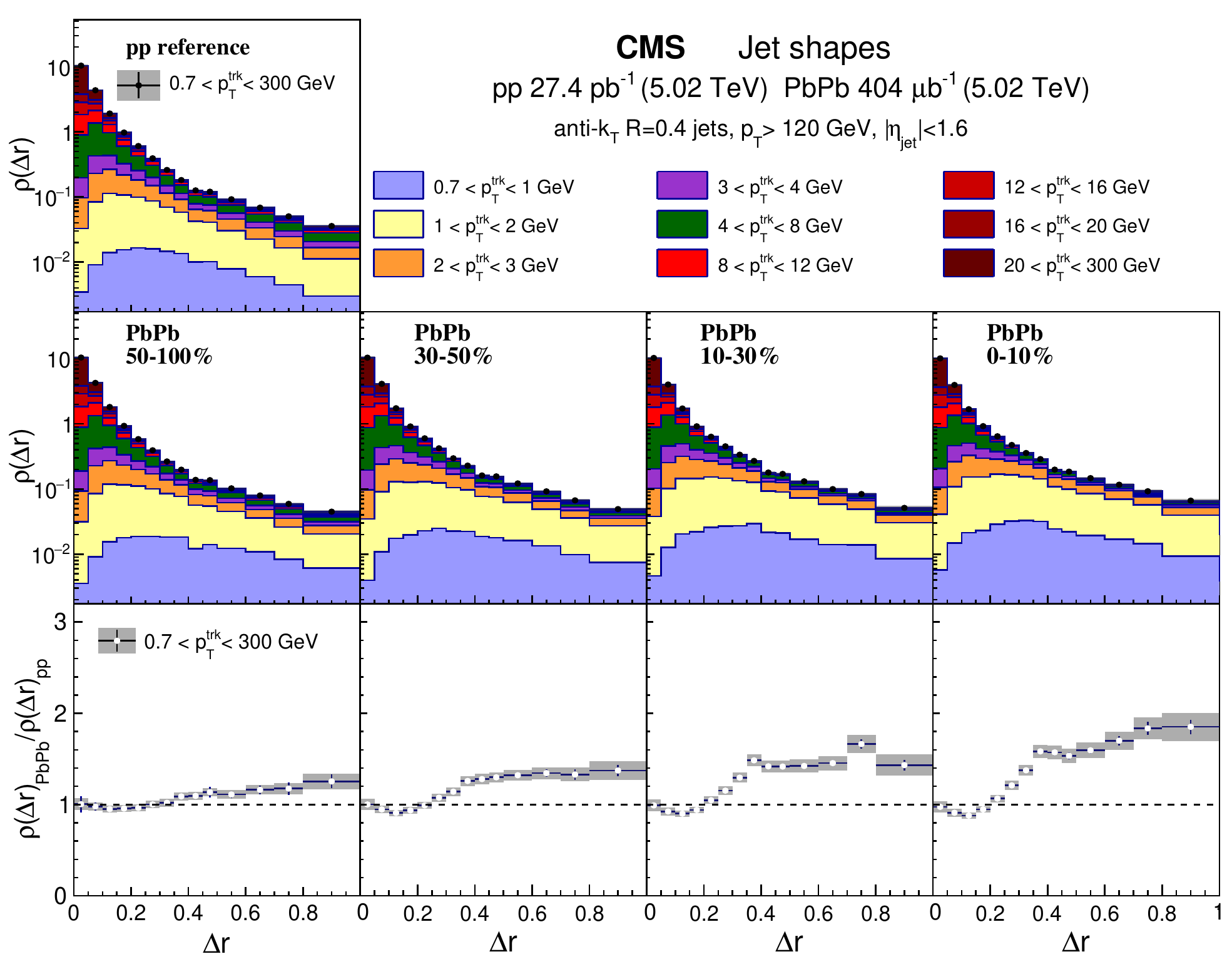}
         \caption{The jet shape $\rho(\Deltar)$ in $\Pp\Pp$ (top left) and PbPb (middle row) collisions. The PbPb results are shown for  different centrality regions. The bottom row shows the ratio between the PbPb and $\Pp\Pp$ data  for the inclusive range $0.7<\pttrk<300\GeV$. The shaded band in the ratio plots shows the total systematic uncertainty.}
       \label{fig:jetshape_dr_stacked}
 \end{figure}

The analogous results for $\rho(\Deltar)$ are shown in Fig.~\ref{fig:jetshape_dr_stacked}. The ratios of the PbPb to the $\Pp\Pp$ data, presented in the bottom row of Fig.~\ref{fig:jetshape_dr_stacked}, are shown for  the inclusive range $0.7<\pttrk<300$\GeV. A redistribution of energy from small to large angles relative to the jet axis is evident from these data, as seen (for example) from the dip and then rise in the PbPb/$\Pp\Pp$ ratio distributions as $\Deltar$ increases. The effect is more pronounced in central events. From Fig.~\ref{fig:jetshape_dr_stacked}, we conclude that this large enhancement at large $\Deltar$  is dominated by the contribution from low-\pttrk tracks, has pronounced centrality dependence, and, as shown in Fig.~\ref{fig:yield_integral}, is slightly larger at higher center-of-mass energy.

Note that for the 2.76\TeV data, shown in Fig.~2 of Ref.~\cite{HIN_2015_011}, the PbPb/$\Pp\Pp$ jet shape ratios for leading jets at large $\Deltar$ are larger than those shown for inclusive jets from 5.02\GeV in Fig.~\ref{fig:jetshape_dr_stacked} of this manuscript. This is mainly because the jet shape in the 2.76\TeV $\Pp\Pp$ reference data falls more steeply as a function of $\Deltar$ than at 5.02\TeV, resulting in a larger value for the 2.76\TeV ratio even though the magnitude of the modification effects is similar at the two energies. This difference in the shape of the two $\Pp\Pp$ reference samples is also present in the \PYTHIA simulation, where the effect is due to differences in the relative fraction of quark and gluon jets at the two  energies.

\section{Summary}
\label{sec:Summary}

In this paper, measurements are presented of the modifications to charged-particle track yields and jet shapes in PbPb collisions with respect to $\Pp\Pp$ collisions using data collected with the CMS detector at the LHC at a nucleon-nucleon center-of-mass energy of $\sqrtsNN = 5.02\TeV$.
The correlations of charged particles having transverse momentum $\pttrk > 0.7$\GeV and pseudorapidity $\abs{\eta}<2.4$ with the jet axis of jets having $\pt > 120$\GeV and $\abs{\eta}<1.6$ are studied. Charged particle yields associated with jets are shown as a function of relative angular distance $\Deltar = \sqrt{\smash[b]{(\Delta\eta)^{2}+(\Delta\phi)^{2}}}$ from the jet axis, as well as individually in $\Delta\eta$ and $\Delta\phi$.  In these studies, a strong enhancement of tracks with $\pttrk<3$\GeV, extending to large angles, is found in PbPb collisions with respect to $\Pp\Pp$ collisions.
This low \pt excess remains correlated with the jet axis, but the distribution is broader in PbPb than that observed in the corresponding \pt bin of $\Pp\Pp$ collisions, which could indicate additional in-medium gluon radiation and/or a medium backreaction, \ie, a wake-like response of the QGP to the propagating parton. For PbPb events with centrality 0--10\%, the correlated yield is increased relative to $\Pp\Pp$ collisions at low \pt by up to a factor of two.

In addition to these charged-particle yields, we examine the jet transverse momentum profile $\Rhodr$ and the jet shape $\rho(\Deltar)$ variables, defined using the distribution of charged-particle tracks in annular rings around the jet axis, with each particle weighted by its \pttrk value.
A redistribution of energy from small to large angles from the jet axis is observed for PbPb relative to $\Pp\Pp$ events, with the most pronounced effect seen for central collisions. The energy flow within the jet is modified by shifting the momentum away from the jet axis out to large relative angular distances such that in central PbPb collisions, the ratio of the jet shapes in PbPb to $\Pp\Pp$ collisions approaches a value of two. These measurements provide  a comprehensive picture of the modifications of the parton shower evolution and the quark-gluon plasma response to the propagating jets in 5\TeV PbPb collisions.

\begin{acknowledgments}
We congratulate our colleagues in the CERN accelerator departments for the excellent performance of the LHC and thank the technical and administrative staffs at CERN and at other CMS institutes for their contributions to the success of the CMS effort. In addition, we gratefully acknowledge the computing centres and personnel of the Worldwide LHC Computing Grid for delivering so effectively the computing infrastructure essential to our analyses. Finally, we acknowledge the enduring support for the construction and operation of the LHC and the CMS detector provided by the following funding agencies: BMWFW and FWF (Austria); FNRS and FWO (Belgium); CNPq, CAPES, FAPERJ, and FAPESP (Brazil); MES (Bulgaria); CERN; CAS, MoST, and NSFC (China); COLCIENCIAS (Colombia); MSES and CSF (Croatia); RPF (Cyprus); SENESCYT (Ecuador); MoER, ERC IUT, and ERDF (Estonia); Academy of Finland, MEC, and HIP (Finland); CEA and CNRS/IN2P3 (France); BMBF, DFG, and HGF (Germany); GSRT (Greece); OTKA and NIH (Hungary); DAE and DST (India); IPM (Iran); SFI (Ireland); INFN (Italy); MSIP and NRF (Republic of Korea); LAS (Lithuania); MOE and UM (Malaysia); BUAP, CINVESTAV, CONACYT, LNS, SEP, and UASLP-FAI (Mexico); MBIE (New Zealand); PAEC (Pakistan); MSHE and NSC (Poland); FCT (Portugal); JINR (Dubna); MON, RosAtom, RAS, RFBR and RAEP (Russia); MESTD (Serbia); SEIDI, CPAN, PCTI and FEDER (Spain); Swiss Funding Agencies (Switzerland); MST (Taipei); ThEPCenter, IPST, STAR, and NSTDA (Thailand); TUBITAK and TAEK (Turkey); NASU and SFFR (Ukraine); STFC (United Kingdom); DOE and NSF (USA).

\hyphenation{Rachada-pisek} Individuals have received support from the Marie-Curie programme and the European Research Council and Horizon 2020 Grant, contract No. 675440 (European Union); the Leventis Foundation; the A. P. Sloan Foundation; the Alexander von Humboldt Foundation; the Belgian Federal Science Policy Office; the Fonds pour la Formation \`a la Recherche dans l'Industrie et dans l'Agriculture (FRIA-Belgium); the Agentschap voor Innovatie door Wetenschap en Technologie (IWT-Belgium); the Ministry of Education, Youth and Sports (MEYS) of the Czech Republic; the Council of Science and Industrial Research, India; the HOMING PLUS programme of the Foundation for Polish Science, cofinanced from European Union, Regional Development Fund, the Mobility Plus programme of the Ministry of Science and Higher Education, the National Science Center (Poland), contracts Harmonia 2014/14/M/ST2/00428, Opus 2014/13/B/ST2/02543, 2014/15/B/ST2/03998, and 2015/19/B/ST2/02861, Sonata-bis 2012/07/E/ST2/01406; the National Priorities Research Program by Qatar National Research Fund; the Programa Severo Ochoa del Principado de Asturias; the Thalis and Aristeia programmes cofinanced by EU-ESF and the Greek NSRF; the Rachadapisek Sompot Fund for Postdoctoral Fellowship, Chulalongkorn University and the Chulalongkorn Academic into Its 2nd Century Project Advancement Project (Thailand); the Welch Foundation, contract C-1845; and the Weston Havens Foundation (USA).
\end{acknowledgments}

\bibliography{auto_generated}

\cleardoublepage \appendix\section{The CMS Collaboration \label{app:collab}}\begin{sloppypar}\hyphenpenalty=5000\widowpenalty=500\clubpenalty=5000\textbf{Yerevan Physics Institute,  Yerevan,  Armenia}\\*[0pt]
A.M.~Sirunyan, A.~Tumasyan
\vskip\cmsinstskip
\textbf{Institut f\"{u}r Hochenergiephysik,  Wien,  Austria}\\*[0pt]
W.~Adam, F.~Ambrogi, E.~Asilar, T.~Bergauer, J.~Brandstetter, E.~Brondolin, M.~Dragicevic, J.~Er\"{o}, A.~Escalante Del Valle, M.~Flechl, M.~Friedl, R.~Fr\"{u}hwirth\cmsAuthorMark{1}, V.M.~Ghete, J.~Hrubec, M.~Jeitler\cmsAuthorMark{1}, N.~Krammer, I.~Kr\"{a}tschmer, D.~Liko, T.~Madlener, I.~Mikulec, N.~Rad, H.~Rohringer, J.~Schieck\cmsAuthorMark{1}, R.~Sch\"{o}fbeck, M.~Spanring, D.~Spitzbart, A.~Taurok, W.~Waltenberger, J.~Wittmann, C.-E.~Wulz\cmsAuthorMark{1}, M.~Zarucki
% ...
\vskip\cmsinstskip
% ...
1:~~Also at Vienna University of Technology, Vienna, Austria\\
% ...

\end{sloppypar}
\end{document}